\documentclass[journal,onecolumn]{IEEEtran}
\ifCLASSINFOpdf
\else
\fi

\usepackage{caption} 
\usepackage{graphicx}
\usepackage{comment}
\graphicspath{{/home/anamika/git_repo_master/Others/}}
\usepackage{array}
\usepackage[outdir = ./]{epstopdf}

\usepackage{array}
\newcolumntype{P}[1]{>{\raggedright\arraybackslash}p{#1}}

\IEEEoverridecommandlockouts

\IEEEpubid{\begin{minipage}{\textwidth}\ \\[12pt]
(A preliminary version of this paper is available in the proceedings of OpenSym 2018.)
 \end{minipage}}

\begin{document}
%
\title{Capturing Knowledge Triggering in Collaborative Settings}
%
%
%
\author{Anamika Chhabra, \textit{IIT Ropar, India} and S. R. Sudarshan Iyengar, \textit{IIT Ropar, India}}

\maketitle

\begin{abstract}


In collaborative knowledge building settings, the existing knowledge in the system is perceived to set stage for the manifestation of more knowledge, termed as the phenomenon of \textit{triggering}.
Although the literature points to a few theories supporting the existence of this phenomenon, these have never been validated in real collaborative environments, thus questioning their general prevalence. In this work, we provide a mechanized way to observe the presence of triggering in knowledge building environments. We implement the method on the most-edited articles of Wikipedia and show how the existing \textit{factoids} lead to the inclusion of more factoids in these articles. The proposed technique may further be used in other collaborative knowledge building settings as well. The insights obtained from the study will help the portal designers to build portals enabling optimal triggering.
\end{abstract}


%
\IEEEpeerreviewmaketitle

\section{Introduction}

Human knowledge is perceived to evolve with time through the creation of new artifacts of knowledge. 
An important thing about the evolution of knowledge is that the development of new artifacts depends on the existing ones~\cite[Pg. 86]{ogburn1922inventions}. 
This belief is also supported by \textit{Constructivist learning theories}~\cite{anderson1998online} as well as other theories on \textit{Genetic Epistemology}~\cite{piaget2013construction, cooper1993paradigm, ertmer1993behaviorism, vygotskie1978mind, luhmann1995social, piaget1976piaget}. As per these theories, how we construct knowledge depends on what is already known. Understanding this evolutionary process has been of interest to the researchers from a long time~\cite{minsky1977frame, fisher1985information}. Many of these works have pointed to the phenomenon of \textit{triggering} to be responsible for the creation of new knowledge. Here, triggering is a procedure by which an idea or a comment spearheads the generation of another idea or thought~\cite{norman1981categorization}. Although this phenomenon has been mentioned in different texts in diverse ways, yet they all refer to the same underlying idea. For instance, classical theories suggest that in a social system such as a collaborative knowledge building system, people add more content due to the \textit{cognitive conflicts}~\cite{luhmann1995social} or \textit{perturbations}~\cite{piaget1976piaget}. These conflicts arise when they see content that is not complete or does not match with what is there in their cognitive systems (i.e., minds) already. 

\begin{figure}[!htpb]
\centering
	\includegraphics[scale=0.45]{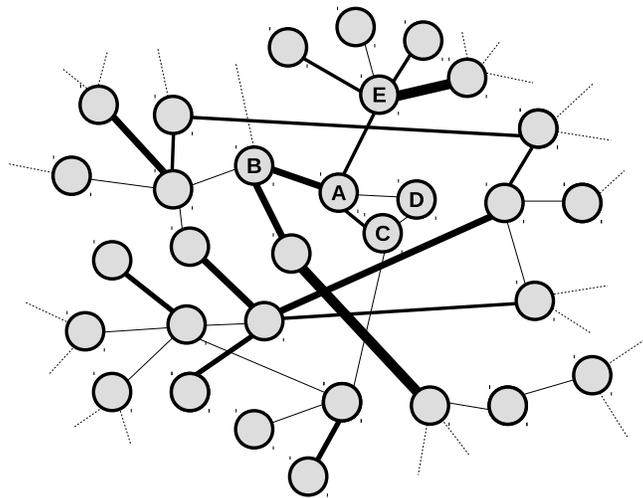}
\caption {Triggering Network: The nodes represent the concepts and a link between two nodes shows that the two concepts are related to each other. The thickness of the edges represents the strength of association between the concepts.}
\label{fig:t-net}
\end{figure}

A few other works point to theories that support the existence of an underlying network among the pieces of knowledge concerning a knowledge artifact~\cite{minsky1977frame, fisher1985information, norman1981categorization}. For example, it is perceived that knowledge is organized into frames and each of these frames possesses a particular concept~\cite{minsky1977frame}. These frames may be of varying sizes, and those that are related to each other are linked together in the network \cite{norman1981categorization}. Therefore, when a frame is triggered, the other frames that are linked to it are also likely to be triggered~\cite[p. 55]{fisher1985information}. These frames may be linked sequentially or in any non-linear fashion. One can imagine a forest of nodes where each node is a knowledge frame, and the attached frames form the connected components in the forest. Figure~\ref{fig:t-net} shows an example of the underlying network of the knowledge frames (concepts) which are shown as nodes and a link between two frames depicts that they are associated with each other. Further, these frames are connected by \textit{condition-action} rules, that determine which frames to trigger next~\cite{fisher1985information}. When the triggering conditions for a frame are met, that frame is brought into the system. Figure~\ref{fig:t-net} captures this by the thickness of the edges that represents the strength of association between the nodes. As an instance, the concept `A' is associated with four more concepts, namely `B', `C', `D' and `E', where A's association with `B' is more than that with `C' and `E', which is further more than that with `D'. Hence when `A' gets introduced, the chances of inclusion of `B', `C', `D' and `E' also increase based on the strength of their edges. This phenomenon leads to a ubiquitous and self-regulating phenomenon of the existing knowledge frames leading to the inclusion of more knowledge into the system, making it an \textit{autopoietic system}.\footnote{An autopoietic system is one in which ``subsequent
operations build on the results of the preceding operations''~\cite{cress2008systemic}.}

A limitation of the existing works on understanding the evolution of knowledge is that they have suggested theories that have not been validated with real-world data. An inability to get access to the right kind of data has mainly resulted in only theoretical work pursued in this direction. However, in the recent past, due to the advancements of tools enabling large-scale collaboration for building knowledge, it is possible to 
acquire the required underlying data. This is due to the fact that these tools store all the footprints of the knowledge built through them in digital format. 

In this study, we use Wikipedia which is one of the most successful portals that enables building knowledge with a combined effort of a large group of users.  
On this portal, the content available in any given article does not reach its eventual state in a single step~\cite{nunes2008wikichanges}. Rather, a few \textit{factoids}\footnote{A factoid may refer to a standalone piece of information about the topic of the article.} get added in the beginning, which are then read by other users instigating them to add the connected factoids and so on~\cite{de2017grand}. 
These factoids form a meaningful network of associations. So when a new factoid is added, it increases the likelihood of addition of more factoids that are linked to it. For instance, the addition of information on `Vegetarianism' in the article on `India' increases the chances of inclusion of information about `Jain' and `Brahmin' community which strictly follow vegetarianism. In other words, the existing content of the articles triggers the users to contribute more content leading to the evolution of the articles through subsequent revisions. It should be noted that although Wikipedia is a knowledge accumulation platform, yet it emulates the knowledge construction process in multiple ways~\cite{chhabra2017does}. Further, a large scale collaboration as well as elaborate digital footprints of the successive changes made to its articles makes it an ideal playground to understand the triggering phenomenon. We, therefore, use data from Wikipedia to understand the dynamics of evolution of knowledge. We propose a technique to automatically capture the triggering phenomenon among the important factoids of its articles. We further show in-depth observations made on a few most edited articles which provide evidence towards the process of existing content leading to the inclusion of new content. 

The development of mechanized ways to observe the evolution of a collaborative piece of knowledge may pave the way for advancements in the fundamental research on knowledge building. This will further lead to better mechanism design of the collaborative tools for building knowledge.

\section{Related Work}

Past research has attempted developing models that mimic the triggering phenomenon. However, these models have focused mainly on the growth properties of the knowledge rather than exploring how the existing knowledge frames steer the inclusion of more knowledge.


Some of these models are based on \textit{Polya's Urn Model}~\cite{dosi1994generalized} and its extensions~\cite{mahmoud2008polya, marengo2016arrival}  where the knowledge units are represented as colored balls in an urn. In the basic model, one ball is drawn from the urn uniformly at random and its color is observed. This ball along with an extra ball having the same color is then inserted into the urn. This process is repeated and the growth of balls of different colors in the urn is observed. It is understandable that in such a model, an observed ball (color) is more likely to be observed again.Tria et al.~\cite{tria2014dynamics} present a mathematical model to emulate the occurrence of a new invention. The model is also a generalization of the Polya's Urn model~\cite{mahmoud2008polya} and is based on the idea that the space of the existing novelties expands as a new invention occurs. It uses the concept of \textit{adjacent possible} introduced by Kauffman~\cite{kauffman1996investigations}, where the adjacent possible may contain all those concepts that are one step away from the existing concepts~\cite{tebbe2011good}. The authors show that the rate of occurrence of novelties follows \textit{Heap's Law}~\cite{heaps1978information}. As per this law, the growth rate of knowledge reduces wih time. The same was then verified using the data taken from various sources such as Wikipedia, an annotation system and an online music catalog. The Wikipedia dataset used by the authors contained a collection of Wikipages and the first edit by a user to a wikipage was considered equivalent to an invention. This model was further extended by Loreto et al.~\cite{loreto2016dynamics} where the authors provided a class of probabilistic models using \textit{Simon's model}. 
In a recent work, Iacopini et al. \cite{iacopini2018network} modeled the dynamics of innovation processes using \textit{Edge Reinforced Random Walks}~\cite{keane2000edge} on the network of ideas. Edge Reinforced Random walk is one in which the weights of the edges are incremented as they are visited, thereby, changing the weights of the edges in the network dynamically. The authors kept the probability of visiting an edge to be directly proportional to the weight of the edge. They showed that the rate of increase of innovations through their model follows Heaps law, which has been shown to exist for such settings by past literature. In another work, the knowledge networks of questions and answers were studied by Miroslav et al.~\cite{andjelkovic2016topology}. 
Using the concept of triggering from the classical cognitive theories, Chhabra et al.~\cite{chhabra2015skillset} developed a mathematical model that computes the knowledge produced in a system due to the effect of triggering. The model uses the concept of diversity in activity selection behavior of users in a collaborative environment~\cite{chhabra2015presence, chhabra2015characteristic}.

All these models have mainly focused on either finding the rate at which the inventions occur or some other statistical property of the process rather than understanding how the knowledge evolves through the process of triggering. 
To the best of our knowledge, how the existing knowledge sets the stage for the manifestation of more knowledge has not been explored using real-world data so far.

\section{Dataset}
The data set\footnote{Collected in November 2017.} used for the analysis contains the entire revision history of the top 100 most \textit{edited} articles on Wikipedia. The rationale behind choosing the most-edited articles is that the phenomenon of triggering may be better understood by analyzing the articles that have accumulated a large number of edits as compared to the articles with a comparatively smaller revision history. The data is in XML format and contains details such as username or IP address (if the user was anonymous), user Id, revision Id, the entire content of the article after the edit that lead to that particular revision, timestamp of the revision, the article size in bytes etc. We specifically did not consider `list' articles which mostly contain links to other Wikipedia articles on some topic, such as `List of Programs Broadcast by GMA Network' and `List of Impact Wrestling Personnel'. This is due to the fact that these articles are built in a slightly different manner as compared to the rest of the articles where the editing happens at word-level. The data set contains articles on a wide variety of topics ranging from people such as `George W. Bush', `Britney Spears', `Beyonce' to countries such as `India', `United States', `United Kingdom' to general topics such as `Same-sex Marriage', `September 11 Attacks' and `Christianity'. Except for one article viz. `Syrian Civil War' that was created in 2011, all the articles were created before 2006, with precisely 65 of them created in 2001. 
It was interesting to observe that despite a large number of edits, many of the articles belonged to B and C quality grades. Similarly, a few articles with lesser number of edits could also achieve a better quality grade, indicating that a large number of edits, although improve the quality, however, may not directly imply the quality of the content. For example, with only 15621 edits, the article `Hillary Clinton' achieved a `Featured Article (FA)' status, whereas despite 34625 edits, `Wikipedia' article falls under `B' quality grade class.

\section{Article Evolution: Tracking the Factoids}

A Wikipedia article is always in-flux, i.e., it keeps changing with time as new units of information, i.e., factoids keep getting added. The introduction of these factoids is what leads to the evolution of the article. Therefore, it is required that out of the entire content of the article, these factoids be identified. A manual assessment of an article may provide a clue to a domain expert about what may be considered as factoids out of the entire content of the article. However, in order to capture the evolution through automated techniques, it is essential to devise some measure to identify these important units of information.
\begin{figure*}[!htpb]
\centering
	\includegraphics[scale=0.67]{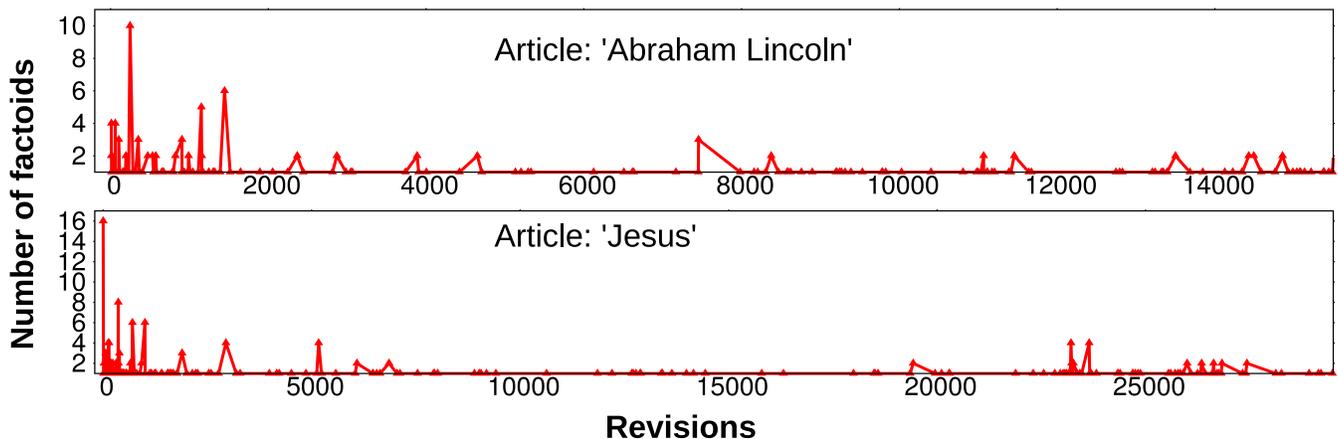}
\caption{Number of factoids introduced in each revision for the articles `Abraham Lincoln' (Top) and `Jesus' (Bottom). The revisions on the x-axis are in the order of their timestamps. A lot of factoids are added in the initial few revisions. }
\label{fig:terms_per_rev_AL_J}
\end{figure*}

Every Wikipedia article contains a number of \textit{Internal links} which point to other Wikipedia articles. These links may contain either a single word such as `Bible' or a phrase such as `Second Samoan Civil War'. We posit that an article is created for a given word or phrase if it is important. Therefore, in our analysis, we use the internal links of an article as a proxy for its important terms or factoids. Moreover, out of all the terms, those that stay till the end, i.e. remain in the final version of the article are even more important. Keeping track of the time of introduction and the inter-dependency among these factoids may provide insights into the evolution of the article. Further, the frequency with which factoids are introduced may also help in revealing which phase an article is currently in. For example, it is understandable that when an article is newly created, the frequency might be greater as compared to the later phases.

We were interested to observe how the articles in the data set evolved to reach their final state. For that, we gathered all the factoids present in the latest version of the articles. We then recorded the revisions where these factoids were first introduced. Figure~\ref{fig:terms_per_rev_AL_J} shows the number of factoids introduced with respect to their revision numbers for two of the articles: `Abraham Lincoln' and `Jesus'. These articles were arbitrarily chosen out of the data set and a similar pattern was observed for the other articles in the dataset as well. The revisions on the X-axis are in the order of their timestamps. As expected, a large number of factoids are added in the first few revisions, whereas the frequency of addition of new factoids reduces in the later revisions.


\begin{figure}[!htpb]
\centering
	\includegraphics[scale=0.62]{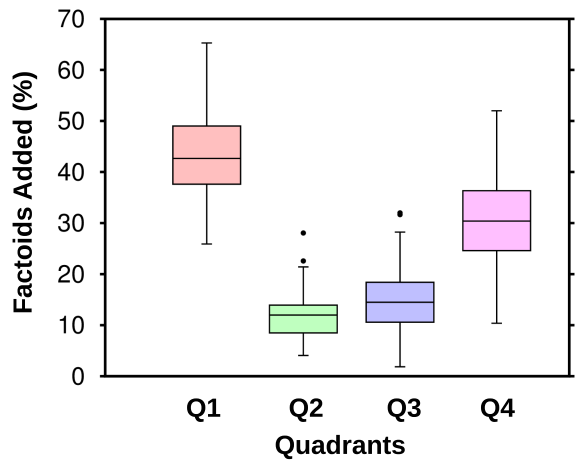}
\caption{Percentage of factoids added in each quadrant, where the quadrants contain the revisions sorted as per their timestamps. Maximum number of terms were added in the first quadrant.}
\label{fig:revid_numterms_box}
\end{figure}

Further, in order to determine an aggregate behavior for all the articles in the data set, we divided the lifespan of each article into four quadrants- $Q_1$, $Q_2$, $Q_3$ and $Q_4$ respectively. We then computed the fraction of factoids that were introduced in each quadrant. Figure~\ref{fig:revid_numterms_box} shows the overall percentage of factoids introduced across all the articles. Intuitively, when the article is in its inception, there is more scope of inclusion of new pieces of information, hence $Q_1$ has the maximum fraction of factoids. The number of factoids introduced in $Q_2$ and $Q_3$ were found to be comparatively lesser. However, $Q_4$ showed an increase in the number of factoids as compared to $Q_2$ and $Q_3$. We feel that around this period, the articles were competing for the status of a `Featured/Good Article' leading to a relative increase in the addition of new pieces of information.

\section{Measuring Triggering Among Factoids}
Triggering is a cognitive and sometimes an individual-specific phenomenon. Automatically perceiving the presence of triggering among the factoids, as well as discerning what may have instigated a user to add their own content, is a challenging task. The non-trivial nature of the analysis is what has led to only theoretical evidence of the presence of triggering phenomenon in existing literature. 

We propose the use of \textit{Normalized Google Distance} (NGD)~\cite{cilibrasi2007google} to measure triggering among factoids and show that it may help in automatically measuring triggering in a collaborative environment to a reasonably good extent. It should further be noted that there are a few other association measures such as network distance or semantic distance techniques such as \textit{word2vec}, however, these methods may not serve our purpose. Network distance may not help because in our analysis, factoids represent Wikipedia articles and the distance between any two articles on Wikipedia network may be able to provide values in a very small range only. This is because the Wikipedia network follows \textit{small-world phenomenon}~\cite{watts1998collective} which leads to a very small average distance between any two nodes on the network\footnote{It has been observed that Wikipedia network is a classic example of a small-world network which is so densely hyperlinked that on an average, it takes only 4.5 clicks to go from one article to another~\cite{dolan2008wikismallworld}.}. Further, in some cases, the presence of a link between any two articles on Wikipedia may be the result of a few parameters specific to Wikipedia and its policies, leading to an association value which may not be universal. On the other hand, semantic distance measures such as word2vec provide association between words, whereas our analysis requires finding association between factoids, which may be phrases having a collection of words, both nouns, and pronouns. For instance, such measures will not be able to provide any details about the relation between phrases such as `Battle of Long Island' and `New-York Historical Society'. 

%
%
%

NGD is a sort of `crowdsourced' way of computing the semantic similarity between two words or phrases. It is based on computing the number of hits that are returned by the Google search of these phrases. It exploits the idea that the phrases which are semantically similar will be found together in more number of web pages as compared to those that are not quite related to each other. The formula of NGD is given by:

\begin{equation}\label{NGD}
NGD(a,b) = \frac{\max\{\log{h(a)}, \log{h(b)}\}- \log{h(a,b)}}{\log{N} - \min\{\log{h(a)}, \log{h(b)}\}}
\end{equation}
where $a$ and $b$ are the phrases between which the semantic distance has to be computed. Here, $h(a)$, $h(b)$ and $h(a,b)$ are the number of hits returned by the Google search on the phrases $a$, $b$ and $a$, $b$ together, respectively. Further, $N$ is the total number of web pages examined by the Google query, multiplied by the average number of words on any web page. An estimate of the total number of web pages is found by searching for a word such as `the', that is found on almost every page, which at the time of the study came out to be 25,27,00,00,000. Further, for our analysis, we took the average number of words on any web page to be 1,000. Although the value of NGD between two phrases $a$ and $b$ can vary from $0$ to $\infty$, however, if it is greater than 1, $a$ and $b$ are considered to be reasonably dissimilar. A value of $0$ for the metric indicates that the phrases are very related and \textit{always} occur together. Next, we explain how we used NGD to analyze triggering in Wikipedia articles.

\begin{figure}[!htpb]
\centering
	\includegraphics[scale=0.42]{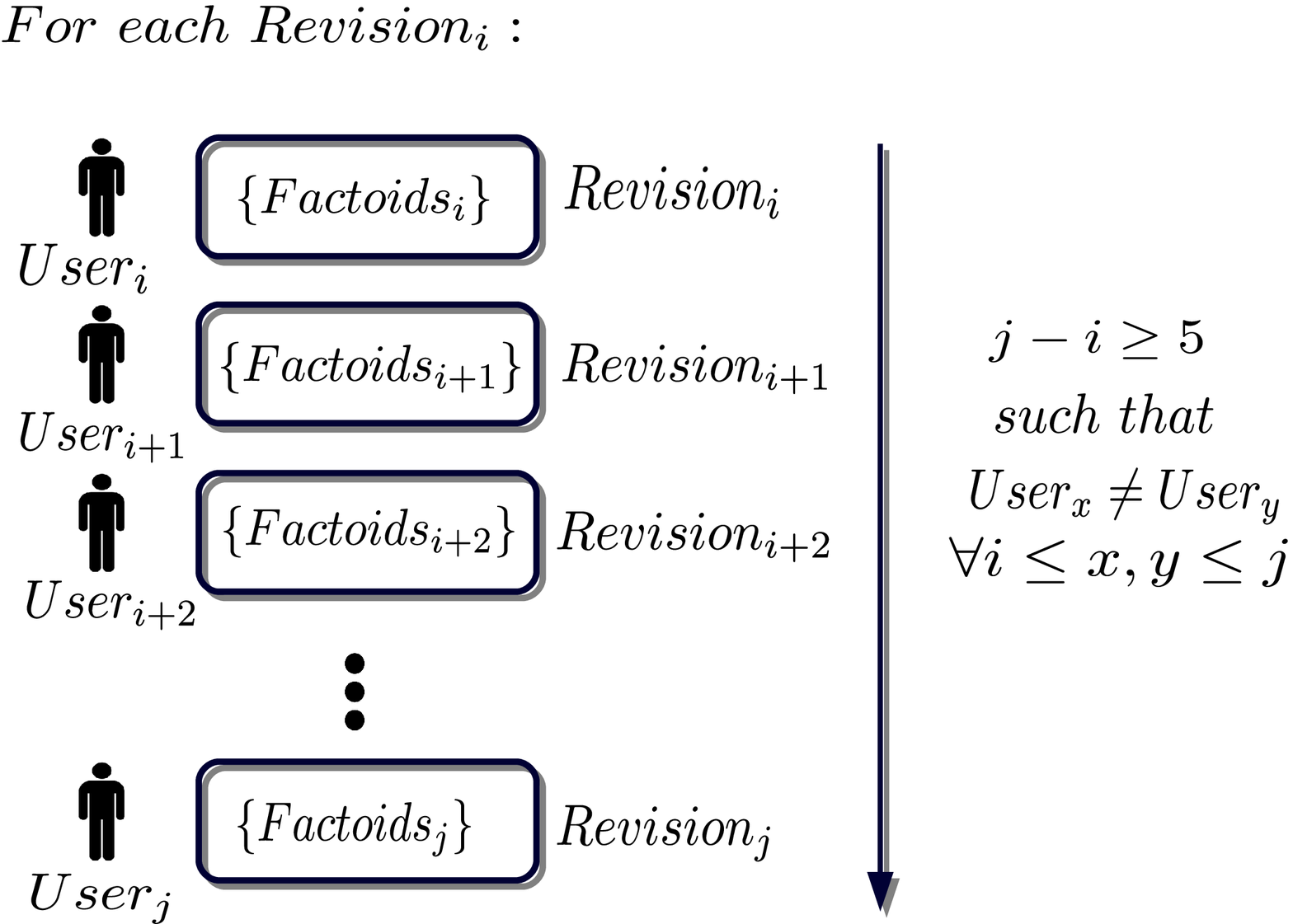}
\caption{Factoid set and user in each revision were recorded. For each revision, the contribution made by the next five users was analyzed.}
\label{method}
\end{figure}

For each Wikipedia article, we first created a list of all the internal links, i.e., factoids that were present in the final version of the article. It should be noted that in this analysis, we considered only the factoids present in the final version, but for a more comprehensive analysis, the factoids introduced in all the revisions - which includes those that got extinct and did not make it to the final version - may also be considered. Subsequently, for every revision, we prepared a list of factoids that were added in that revision. For each revision, the user id of the user who made the revision was also recorded. The next step was to prepare a list of lists, which we named as RFFR, whose each member list was of the following form:\\
\begin{center}
[$Revision_i$, $Factoids_i$, $Factoids_j$, $Revision_j$]
\end{center}
where $Factoids_i$ and $Factoids_j$ were the set of factoids added in $Revision_i$ and $Revision_j$ respectively. Further, $Revision_j$ was among the next subsequent revisions after $Revision_i$ such that the user $User_i$ of $Revision_i$ was not the same as the user $User_j$ of $Revision_j$. Essentially, for each revision $r$, we checked the subsequent \textit{few}\footnote{In the current analysis, we considered the subsequent five revisions such that the users of these revisions were different. In the cases where the same user made the next revision, we considered more revisions accordingly such that we analyze the revisions by at least next five users.} revisions such that the users of these revisions were not the same as the user of revision $r$ (See Figure~\ref{method}). This is because, in this analysis, we aimed to analyze the triggered factoids in a limited number of subsequent revisions, considering that the users get instigated on seeing the recent changes and hence take action. This assumption is particularly true for all those users who are `watchers' or `administrators' that get notified of any changes made to the article. However, the analysis may be extended to analyze more revisions. From the list RFFR, we then removed all those rows, where either $Factoid_i$ or $Factoid_j$ was empty. The intuition is that if the set $Factoids_i$ has some terms, whereas $Factoids_j$ is empty, this means that the terms in $Factoids_i$ were not able to trigger any more terms in the subsequent revisions. On the other hand, if the set $Factoids_i$ is empty, whereas $Factoids_j$ has a few terms, it may indicate that the terms in $Factoids_j$ are independent and were not triggered by the terms in the set $Factoids_i$. We removed all such cases to focus the analysis on the triggered factoids only. However, finding and analyzing the independent factoids along with the triggered ones can be a useful extension. 
As the next step, since we wanted to check which factoid in a revision is highly likely to lead to which other factoids in the subsequent revisions, we recorded all the possible \textit{Factoid-pairs} from 
RFFR. We did this by computing the cross-product of $Factoids_i$ and $Factoids_j$ where $Factoids_{<x>}$ is a set of factoids introduced in the $x^{th}$ revision.
We called the resultant list of lists as RFFR\_cross. 
For example, if RFFR contains a row as : [23356, \{Islam, New York\}, \{Sikhism, Christianity, Hinduism\}, 23357] then RFFR\_cross will contain 2*3 = 6 entries, viz. [23356, Islam, Sikhism, 23357], [23356, Islam, Christianity, 23357], [23356, Islam, Hinduism, 23357], [23356, New York, Sikhism, 23357], [23356, New York, Christianity, 23357] and [23356, New York, Hinduism, 23357]. An analysis of all the factoid-pairs may help us find the most likely pair. Therefore, for every row in RFFR\_cross, we took the factoids of that row and computed the value of NGD as per the Equation~\ref{NGD}. We called the corresponding list as RFFRG where G stands for the NGD value. The next step was to remove from this final list all those rows where the NGD value was above a particular threshold. To compute the threshold, we manually analyzed the articles and observed the extent of association between the factoids vis-a-vis their NGD values. The next section discusses these observations in detail.

\section{Observations}
We computed RFFR and RFFR\_cross for all the articles in the data set. The maximum number of rows in RFFR were 139 for `New York City' article, whereas the maximum number of rows in RFFR\_cross were 596 for `Lionel Messi' article. The average number of rows in RFFR and RFFR\_cross were found to be 39.3 and 146.92 respectively. We then computed the NGD values for each factoid-pair from RFFR\_cross and manually observed the association between the factoids against their NGD values. It was interesting to find a good association between the factoids where the NGD values were less than 0.5, whereas there was very less or no association between the factoids with high values of NGD. To perceive how the existing factoids increase the likelihood of the inclusion of related factoids in the subsequent revisions, we discuss here the results obtained for two of the articles: `India' and `New York City' in detail. The article `India' was chosen due to the domain knowledge of the authors, and `New York City' was chosen given that it had the maximum number of entries in RFFR.

\begin{table}[!htbp]
\centering
\begin{tabular}{l|P{2cm}|l|P{2cm}|l}
\hline
$Revision_i$ & Factoid Added in $Revision_i$ ($a$) & $Revision_j$ & Factoid Added in $Revision_j$ ($b$) & NGD($a$, $b$) \\
\hline
17030 &	NKP Salve Challenger Trophy &	17031 &	2003 Afro-Asian Games &	0.098\\

392 &	Indian Coast Guard	 & 398 &	 Sino-Indian War &	0.099 \\

19235 &	Street cricket	& 19237 &	1936 Summer Olympics	 & 0.122\\

136 &	Sachin Tendulkar &	137 &	gilli-danda &	0.122 \\

17031 &	2003 Afro-Asian Games &	17034 &	Hockey India	& 0.154 \\

52	& Indian subcontinent &	57 &	 Arabian Sea &	0.156 \\

392 &	Indian Army &	398 &	Sino-Indian War &	0.166 \\

136 &	Sachin Tendulkar &	137 &	Kabaddi &	0.166 \\

572 &	Manmohan Singh & 	577 &	Atal Bihari Vajpayee &	0.168 \\

145 &	Mahatma Gandhi &	 148  &	Mohandas Karamchand Gandhi &	 0.216\\

392 &	Indian Air Force &	398	 & Sino-Indian War &	0.235 \\

52 &	 Islam	& 57 & Buddhism &	0.262\\

136 &	Cricket &	137 &	Kabaddi &	0.263 \\

29 &	 Hindi &	 34	& Sanskrit & 	0.275 \\

136 &	Cricket &	137 &	Chess &	0.278 \\

136 &	Sachin Tendulkar	 & 137 &	 Chess &	 0.318 \\

52 & 	Indian subcontinent &	57 &	 Indian Ocean & 	0.322 \\

52 & 	Islam &	57 & 	Christianity &	0.322\\

52 & 	Islam &	57 & 	Sikhism &	0.332 \\

52 &	 Islam	& 57	 & Jainism & 	0.346 \\

244 & Kolkata &	246 &	East India Company &	0.359 \\

52 & 	Islam &	57 & 	Hinduism &	0.360 \\

28 & 	Bengal & 	32 & 	Satyameva Jayate &	0.360 \\

136 &	Cricket &	137 &	gilli-danda	& 0.499 \\

\hline
\end{tabular}
\caption{A few `related' factoid-pairs in the `India' article along with their revision numbers and NGD values which were less than 0.5. Note that the gap between the revisions $Revision_i$ and $Revision_j$ is more in those cases where for a $Revision_i$, some of the subsequent five revisions were made by the same user as that of $revision_i$.}
\label{india-pos-t}
\end{table}

\begin{table}[!htbp]
\centering
\begin{tabular}{l|P{2cm}|l|P{2cm}|l}
\hline
$Revision_i$ & Factoid Added in $Revision_i$ ($a$) & $Revision_j$ & Factoid Added in $Revision_j$ ($b$) & NGD($a$, $b$) \\

\hline
19235 &	Bagepalli &	19237 &	1936 Summer Olympics &	0.568 \\
30 &	 United Kingdom &	32	& Satyameva Jayate & 	0.568\\
13411 &	Current Science	& 13420 &	Cambridge University Press	 & 0.572 \\
30 &	 United Kingdom	& 34 &	Sanskrit &	0.581 \\
28	& Jharkhand &	30	& United Kingdom &	0.665 \\
18643 &	Telecom Regulatory Authority of India &	18667 &	Kedarnath Temple &	0.710 \\
19235 &	Juara &	19237 &	1936 Summer Olympics &	0.974\\
19231 &	UNESCO World Heritage List &	19235 &	Juara &	1.261 \\
\hline
\end{tabular}
\caption{A few `not-so-related' factoid-pairs in the `India' article along with their revision numbers and NGD values, which were more than 0.5.}
\label{india-neg-t}
\end{table}

\begin{table}[h]
\centering
\begin{tabular}{l|P{2cm}|l|P{2cm}|l}
\hline
$Revision_i$ & Factoid Added in $Revision_i$ ($a$) & $Revision_j$ & Factoid Added in $Revision_j$ ($b$) & NGD($a$, $b$) \\
\hline
14999 &	Battle of Long Island &	15021 &	New-York Historical Society	& 0.061 \\

205 &	Skyline &	209 &	Skyscraper &	 0.083 \\

1187 &	Wagner College &	 1189 &	Manhattan College &	0.086 \\

77 &	 Long Island Rail Road &	 79 & LaGuardia Airport & 	0.093 \\

7 &	Staten Island &	11	& Central Park & 0.094 \\

77 & 	Roosevelt Island &	79 & 	LaGuardia Airport &	0.101 \\

355 & Metropolitan Opera & 357 & New York City Public Schools & 0.101 \\

19847 &	New York City Pride March &	19855 &	Riverside Church & 	0.104 \\

15005 & 	Conference House Park &	15021 &	New-York Historical Society & 	0.113 \\

77 & 	Port Authority of New York and New Jersey &	79 &	JFK International Airport	& 0.125 \\

18 & 	New York University	& 19 &	New York Botanical Gardens	& 0.128 \\

2037 &	Throgs Neck Bridge	& 2043 & 	Triborough Bridge & 	0.131\\

77 &	 Port Authority Bus Terminal &	79 & 	LaGuardia Airport &	0.136 \\

1097 &	City park &	1103 &	Battery Park City &	0.148 \\

57 &	 Immigration	 & 65 &	United States Census Bureau &	0.150 \\

77 & People mover &	79 & LaGuardia Airport &	0.152 \\

495 &	News Corporation	 & 497 & 	Television production & 	0.154 \\

71 & 	Bronx Zoo &	77 & 	Long Island Rail Road &	0.155 \\

7 &	George Washington &	13 & 	Columbia University &	0.158 \\

355 &	World War II &	357 &	City University of New York &	0.167 \\

11 & 	Central Park	 & 18 &	Washington Square Park &	0.171\\

7 &	financial center &	12 & 	New York Stock Exchange &	0.179\\

10578 &	General American	 & 10581 &	Italian American &	0.181 \\

77 &	 Long Island Rail Road	& 79 &	JFK International Airport	& 0.181 \\

77 &	Port Authority of New York and New Jersey &	79 &	LaGuardia Airport &	0.188\\

1186 &	Fordham University &	1187 &	Wagner College &	0.203\\

14999 &	Lord Howe &	15021 &	New-York Historical Society &	0.208\\

\hline
\end{tabular}
\caption{A few `related' factoid-pairs in the `New York City' article along with their revision numbers and NGD values which were less than 0.21.}
\label{nyc-pos-t}
\end{table}

\begin{table}[h]
\centering
\begin{tabular}{l|P{2cm}|l|P{2cm}|l}
\hline
$Revision_i$ & Factoid Added in $Revision_i$ ($a$) & $Revision_j$ & Factoid Added in $Revision_j$ ($b$) & NGD($a$, $b$) \\
\hline
395 &	Jerusalem	 & 425 &	stadium &	0.556\\

69 &	 disco &	 71 &	Bronx Zoo &	0.573\\

28 &	 New Netherland &	33 &	 song &	0.579\\

691 &	Metro-North Railroad &	697 &	United Kingdom &	0.582\\

7 &	port &	13	& New York Yankees &	 0.601\\

355 & Brazil &	357 &	Fashion Institute of Technology &	0.634 \\

\hline
\end{tabular}
\caption{A few `not-so-related' factoid-pairs in the `New York City' article along with their revision numbers and NGD values, which were more than 0.5.}
\label{nyc-neg-t}
\end{table}

%
%


The total number of factoid-pairs for `India' and `New York City' articles were found to be 305 and 533 respectively. The NGD values for all these pairs were computed, and the pairs were sorted based on their NGD values. The minimum and maximum NGD values observed for `India' were 0.042 and 1.26 respectively, whereas for `New York City', these were 0.044 and 0.63 respectively. We observed the association between the factoids against their NGD values. It was interesting to observe that the pairs where the NGD values were less, i.e. the top entries of the sorted list, a strong association was found among the factoids. However, as we went down the list, the association between the factoids reduced. In fact, the factoids in the bottom-most rows were found to be having a very high conceptual distance. Tables~\ref{india-pos-t} and \ref{india-neg-t} show a few entries from the top-most rows and the bottom-most rows respectively, out of the sorted list for `India' article. Tables~\ref{nyc-pos-t} and \ref{nyc-neg-t} show the corresponding values for `New York City' article. It can be seen that the inclusion of the terms `Sachin Tendulkar' and `Cricket' led to the introduction of terms `Kabaddi', `Chess' and `Gilli danda' - which are other games played in India - in the very next revision. The inclusion of `Islam' lead to the inclusion of terms `Christianity', `Sikhism' and `Jainism' in the next few revisions. In the `New York City' article, `Wagner College' led to the inclusion of another competitive college of a similar rank, i.e. `Manhattan College'. Similarly, `Metropolitan Opera', which is engaged in deepening student experiences with opera in the schools of New York City triggered `New York City public schools'. It should be noted that in our analysis, we took care of considering only those pairs of factoids where the users of the first and second factoid of the factoid-pair were different. If we observe the entries in Tables~\ref{india-neg-t} and \ref{nyc-neg-t}, we find that the values of NGD higher than 0.5 belong to factoid-pairs having a very less apparent association. Therefore, it may be a good idea to keep the threshold of NGD to be 0.5 in this case and remove the rest of the rows from the obtained factoid-pairs. For India and New York City articles, 52 and 39 pairs respectively were found to be having NGD values more than 0.5. We believe that the choice of a threshold for NGD is dependent on the context. If we wish to find pairs with a very high association only, this threshold may be tuned to a lower value accordingly. 

\begin{figure*}[!htpb]
\centering
	\includegraphics[scale=0.60]{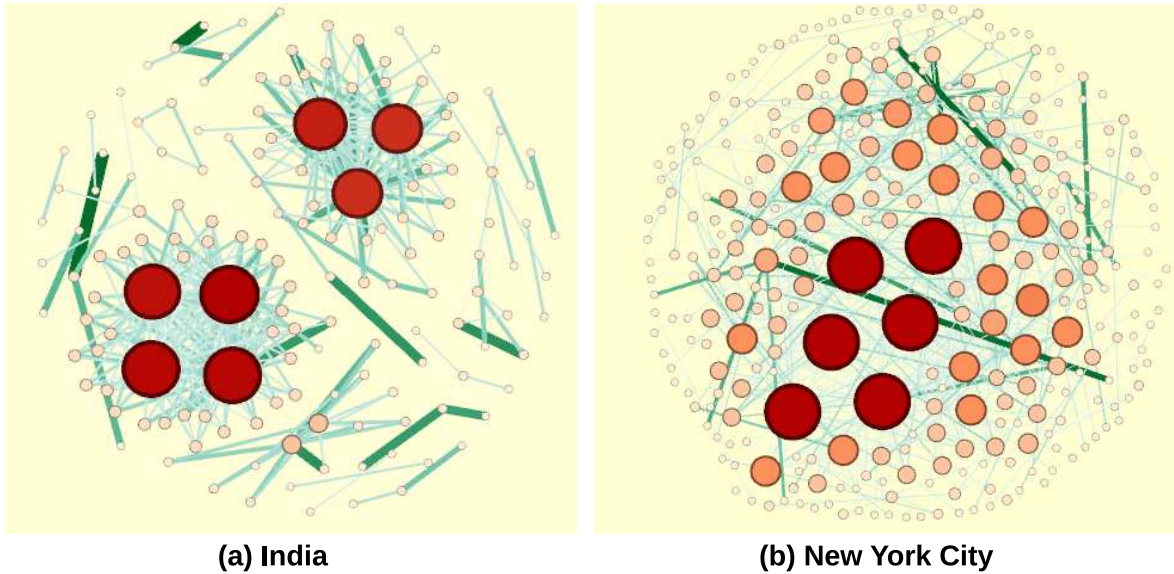}
\caption{Triggering Networks for `India' and `New York City' articles. The darkness of the color of the edges represents the strength of association between the factoids.}
\label{fig:trig-net_I_NYC}
\end{figure*}

We also created the triggering networks for these articles (See Figure~\ref{fig:trig-net_I_NYC}) similar to the one shown in Figure~\ref{fig:t-net} to get an overall picture of triggered terms. Here, the nodes represent the factoids and there is an edge between two factoids if they were added in close-by revisions and one of them had likely led to the inclusion of the other factoid. The size of a node depicts the number of other factoids that a given factoid is connected to. The strength of the association between the factoids is represented by the darkness of the color of the edges, which was computed based on their NGD values. In other words, this strength represents the probability of a factoid getting added to the content of the article, when its connected factoid is already present in the article. This probability is inversely proportional to the value of NGD between the two factoids $a$ and $b$, i.e., 
\begin{center}
Association\_strength(a,b) $\propto$ $\frac{1}{NGD(a,b)}$
\end{center}

Therefore, the computation of NGD values among all the related terms of a given knowledge artifact can help us understand its underlying network and hence its evolution.



\section{Discussion and Future Work}

This paper is a starting step to instigate work in a domain that has remained dormant despite its importance given the extensive usage of crowdsourced portals for building knowledge these days. Triggering is the basis of these portals and an understanding of this phenomenon may help in building interfaces that are able to facilitate optimal triggering. It was interesting to see how the inclusion of a few terms to the articles led to the insertion of more terms in the subsequent revisions. The analysis shows that the introduction of a few key terms acts as milestones for the evolution of the article. When a factoid is added, more knowledge related to that factoid is likely to be added. However, different users get triggered differently, leading to the inclusion of diversified knowledge into the articles. 

To the best of our knowledge, there has not been any automated way to capture triggering among knowledge units in a collaborative setting. Given the difficulty in finding what may have triggered a human mind to add a particular piece of knowledge to an article, it is challenging to devise a foolproof method. Therefore, there may be some limitations of the current analysis as well. For instance, there may be some number of false positives. For example, in our analysis, despite a small value of NGD, it may not be completely guaranteed that the factoids may have been triggered by each other. For example, the NGD between the factoids `American English' and	`Italian American' was 0.079, however, in this case, we are not sure that the former may have necessarily led to the inclusion of the later. At the same time, it should also be noted that given the cognitive nature of the triggering phenomenon, sometimes there may be an indirect and not-so-obvious connection between the factoids for a particular user which is difficult to capture objectively through any automated means.
Nevertheless, one thing that can be clearly considered as a take-away from the proposed analysis is that the usage of NGD method does provide us a probability value where a small NGD value indicates a high probability of one factoid leading to the inclusion of another. In other words, the proposed method does a good job of automating the process of getting the underlying network for an artifact of knowledge that gets built incrementally. 
This information may provide deeper insights into the dynamics of creation of a knowledge artifact. 

There may be various extensions to this work. As an example, the current analysis assumes that triggering happens due to the terms added in the immediate previous revisions only, hence it checks for the triggered terms in the subsequent \textit{few} revisions. The method may be extended to include \textit{all} the successive revisions. 
The decision regarding how many subsequent revisions should be checked also depends on the portal's interface. For example, a portal that sends notifications to its users about any changes made to the content will most likely have triggered terms being added in the close-by revisions as compared to another portal which does not have this facility. Also, the current analysis has been performed only on the terms that remain in the final version. An extended analysis may further be performed on the terms that were present at some point of time in the article, but later got extinct. The underlying network may provide additional insights into the reasons of their extinction. Further, apart from the important \textit{terms}, the same analysis may be performed for all the nouns or all the non-stop-words as well. The properties of the underlying triggering network may also be studied to get deeper insights into the article creation. For instance, this network may provide a clue about demarcating the independent and triggered terms in an article. Additionally, the timestamps of the terms' introduction may provide a directed acyclic graph which may answer many questions regarding the evolution of the articles. The proposed method may also be used in other collaborative settings such as Q\&A portals. 



\section{Conclusion}
In this work, we first suggest a proxy for capturing the important pieces of information in a Wikipedia article. We then show through the analysis performed on some of the most edited articles of Wikipedia that the semantic distance between important terms of a knowledge artifact may help in automatic detection of triggering. We propose the use of Normalized Google Distance as one of the potential measures for computing semantic distance. The analysis may help in understanding the evolution of a piece of knowledge that goes through multiple refinement steps. It may pave way for examining the dynamics of knowledge building on collaborative portals, which has so far remained in the theoretical realms only. This will in turn help in building better crowdsourced portals. 

\bibliographystyle{ieeetran}
\bibliography{/home/anamika/git_repo_master/Others/BIB_PATH/KB_bibliography}

\end{document}